\documentstyle[aps,twocolumn]{revtex}

\begin{document}
\draft
\title{Long-range spatial correlations in the exciton energy distribution in
GaAs/AlGaAs quantum-wells}
\author{Y. Yayon, A. Esser$^{(\ast )}$, M. Rappaport, V. Umansky, H.
Shtrikman and I. Bar-Joseph}
\address{Departement of Condensed Matter Physics, The Weizmann Institute of
Science, Rehovot 76100, Israel\\
(*) AG Halbleitertheorie, Institut fur Physik, Humboldt University Berlin,
Germany}
\maketitle

\begin{abstract}
Variations in the width of a quantum well (QW) are known to be a source of
broadening of the exciton line. Using low temperature near-field optical
microscopy, we have exploited the dependence of exciton energy on well-width
to show that in GaAs QWs, these seemingly random well-width fluctuations
actually exhibit well-defined order --- strong long-range correlations
appearing laterally, in the plane of the QW, as well as vertically, between
QWs grown one on top of the other. We show that these fluctuations are
correlated with the commonly found mound structure on the surface. This is
an intrinsic property of molecular beam epitaxial growth.
\end{abstract}

\pacs{PACS: }

\narrowtext

The confinement of electrons and holes to two dimensions (2D) in
semiconductor quantum wells (QWs) alters the density of states and
interactions, making them an ideal system for studying 2D physics. It is
well known that QW interfaces are not perfect, and some roughness, which
depends on the growth conditions, is always present. As a result, the QW
width is not uniform and the confinement energy fluctuates in the plane.
This gives rise to a disorder potential, which strongly affects the behavior
of particles and quasi-particles in the QW.

This interface disorder potential plays a particularly dominant role in
determining the behavior of QW excitons. These objects are localized in this
potential and their energy spectrum becomes discrete. Indeed, exciton
localization in this disordered system has been a subject of intense
theoretical research, and a variety of spectroscopic techniques were
implemented to study its manifestation in the optical spectrum. Important
insight into this problem came with the application of scanning near-field
optical microscopy (SNOM) to the study of QW photoluminescence (PL). In a
pioneering work, Hess {\it et al.} have directly shown that the far-field
inhomogenously broadened line of the exciton splits into many narrow lines
in the near-field, each line corresponding to a different{\bf \ }exciton
eigenstate of the disordered potential \cite{Hess}. This work was followed
by a number of SNOM measurements that investigated the local properties of
excitons in the spatially fluctuating potential of a QW \cite%
{Gammon,Eytan1,Wu,Wegener,Intonti}. A considerable attention was recently
directed to the statistical properties of the exciton energy levels in this
system \cite{Intonti,Savona}. It was shown that the exciton energy levels in
a disordered narrow QW are not randomly spaced but rather exhibit 'level
repulsion' behavior, manifested by a characteristic spacing between the
energies.

The common assumption in the theoretical studies of exciton localization in
QWs is a disorder potential with only a single short-range correlation
length $\xi $, which is of the order of the exciton Bohr radius $a_{B}$ \cite%
{Zimmermann}. Indeed, previous near-field measurements have focused on
narrow ($\sim 3$ nm) QWs in which single localized excitons can be resolved
and direct insight into the short-range correlations (e.g., extended states,
level repulsion) is obtained \cite{Gammon,Intonti}. In this work we focus on
broad QWs ($10-20$ nm) in which the short-range behavior is averaged by the
SNOM tip, and we study in detail the vertical and lateral long-range order
of the exciton energy distribution. We find that the seemingly random
exciton energy fluctuations exhibit well defined order: strong long-range
correlations appear both laterally, in the plane of the QW, and vertically,
between QWs grown one above the other. We show that this behavior is general
and stems from an intrinsic property of molecular beam epitaxial (MBE)
growth.

The measurements presented in this work were performed on three GaAs QW
samples. Samples 1 and 2 are modulation-doped and have similar structures:
They both have a single 20 nm GaAs QW. In sample 1 a 3.6 nm Si-doped donor
layer is separated from the QW by a 50 nm Al$_{0.35}$Ga$_{0.65}$As spacer
layer and the distance between the QW and the sample surface is 96 nm. In
sample 2 the Si-doped layer has a width of 15 nm, and the spacer is a 60 nm
layer of Al$_{0.35}$Ga$_{0.65}$As. The distance between the QW and the
sample surface is 63 nm. In both samples a $2\times 2$ mm$^{2}$ mesa was
etched, and ohmic contacts were alloyed to the QW layer. A 4.5 nm PdAu
semitransparent gate was evaporated on top of the samples. All the data were
taken at a gate voltage such that the QW is depleted of electrons and the PL
spectrum is dominated by the neutral exciton transition. Sample 3 consists
of four QWs of widths 25, 17, 11 and 5 nm, grown in that order and separated
from each other by 20 nm AlAs layers. The uppermost QW (5 nm) is 50 nm below
the sample surface. The overall growth thickness in samples 1, 2, and 3 are
1.57 $\mu $m, 1.11 $\mu $m, and 1.23 $\mu $m, respectively.

The local measurement of the PL spectrum is performed with a homemade SNOM
operating at 4.2 K \cite{Eytan2}. The sample is illuminated uniformly by a
single-mode fiber and the emitted PL is collected through a tapered
Au-coated etched optical fiber tip \cite{Tips}. The tip collects the PL and
guides it through an optical fiber into a spectrometer, where it is
dispersed onto a liquid nitrogen cooled CCD camera. The overall spectral
resolution is $80$ $\mu $eV. The spatial resolution is determined by the tip
diameter. We have conducted measurements with tips of various diameters in
the range of $0.2-0.3$ $\mu $m .

We begin by studying the lateral distribution of the exciton peak energy $%
E_{X}$\ in the two single QW samples. Figures 1a and 1b are mappings of $%
E_{X}$\ in the two samples, each over an area of $11\times 11$ $\mu $m$^{2}$
($111\times 111$ points). The images are color coded from dark to bright,
corresponding to low and high exciton energies, respectively. The near-field
exciton lineshape in both samples is found to be Gaussian throughout the
measured area, with typical full widths at half maxima of $\sim 0.5$ meV.
The peak of the Gaussian fit allowed $E_{X}$ to be extracted with an
accuracy which is better than our spectral resolution. The near-field
Gaussian lineshape indicates the existence of inhomogenous broadening on the
tip-size scale. The near-field spectrum is therefore a superposition of many
exciton lines, localized in the short-range fluctuations, on a scale smaller
than the tip size.

The two images seem random, with no apparent order in the energy
distribution. It is seen that there is clear distinction between the two: in
sample 1, $E_{X}$\ varies on a significantly shorter length scale than in
sample 2. Examining the fluctuation amplitude another difference is
noticeable: in sample 1 the mean fluctuation is $\sim 0.1$ meV,
significantly smaller than the typical near-field inhomogenous line-width of
the exciton, whereas in sample 2 it is $\sim 0.4$ meV, comparable to the
inhomogenous width. To examine the presence of underlying long-range order
and spatial correlations in these seemingly random energy distributions we
calculated the 2D autocorrelation function, defined as

\begin{equation}
G_{X}(x,y)=<E(x^{\prime },y^{\prime })E(x^{\prime }-x,y^{\prime }-y)>\text{ .%
}
\end{equation}%
Here $<...>$ denotes averaging over all measured points in the scanned area,
and $E(x,y)=$ $E_{X}(x,y)-<E_{X}(x,y)>$. The correlation function averages
out the random behavior and highlights the correlated part of the signal. In
particular, periodic peaks in $G_{X}(x,y)$ indicate the existence of
periodicity in the exciton energy distribution. Figures 1c and 1d show plots
of $G_{X}(x,y)$ for the data of Figs. 1a and 1b, respectively. The plots are
color coded such that bright and dark regions correspond to strong and weak
correlations (large and small $G_{X}$), respectively. Clear symmetry axes
and periodicity are observed in both figures. It is seen that the maxima of $%
G_{X}(x,y)$ in both figures are arranged along lines that match the [100]
and [010] crystallographic orientations \cite{Comment110}. Furthermore,
these maxima exhibit a periodicity of $\sim 1.8$ in sample 1 and $\sim 3$ $%
\mu $m in sample 2. In fact, it is seen that the peaks form a structure that
resembles a {\it cubic lattice}.

The existence of this surprising long-range order along crystallographic
directions clearly indicates that its origin is the structure of the
GaAs/AlGaAs crystal. To understand the relationship between the exciton
energy fluctuations and the crystalline structure we have used the SNOM to
examine the topography of the sample surface $h(x,y)$. We note that the
topography provides us with a measure of the {\it integrated} evolution of
the GaAs/AlGaAs crystal following a growth of $\sim 1$ $\mu $m, while the
exciton energy, which is related to the local well width, gives the {\it %
differential} behavior over 20 nm at the final stage of the growth. Since
the topography is measured in the SNOM simultaneously with the near-field
spectra, one obtains the two signals, $E_{X}(x,y)$ and $h(x,y)$, at exactly
the same location. We find that the samples' surface is covered with mounds
with a typical height of 10 nm (Fig. 2a and 2b). Comparing the force images
to the exciton energy images, one can clearly see the similarity between
them. In Figs. 2c and 2d we show the autocorrelation function $G_{h}(x,y)$
of the topography images for the same areas as in Figs. 1c and 1d. We can
see that the surface topography also exhibits long-range order, which is
similar to that of $E_{X}$. The similarity between $G_{X}(x,y)$ and $%
G_{h}(x,y)$ is especially pronounced in sample 2, where the two functions
exhibit the same symmetry and periodicity. This is demonstrated in Fig. 3,
which shows a cut through the origin of both functions along the [010]
direction. Remarkably, the periodicity is exactly the same for the two lines.

The formation of mounds on surfaces of MBE-grown samples has been a subject
of intensive study during the last decade. However, to the best of our
knowledge, there is no detailed investigation of their influence on the QW
width fluctuations. An important milestone was the finding that their origin
is an intrinsic growth instability, which inhibits downward movement of
adatoms at step edges \cite{Orme,Johnson}. It was shown that the mounds
acquire an elongated shape in the [$1\bar{1}0$] direction due to surface
reconstruction of the As atoms, which terminate the As-rich (001) surface %
\cite{Comment - dangling bonds}. This anisotropy in the mounds' shape is
clearly seen in the topography image of sample 1 (Fig. 2a). As more layers
are added the mounds grow in height and a process of coarsening sets in, in
which small mounds merge with larger ones, and the surface becomes
completely covered with large mounds with a characteristic size and
inclination angle \cite{Orme}. The strong similarity between the long-range
behavior of the exciton energy and the surface topography clearly indicates
that the mound formation plays a dominant role in determining the QW width
fluctuations. In particular, we believe that the coarsening process of the
mound topography is responsible for the quasi-periodic structure that is
observed both in the surface topography and exciton energy. It should be
noted, however, that the exciton image in sample 1 does not show the
anisotropy exhibited by the topography image. This is probably since the
difference in growth along the two crystal directions is not significant for
a thickness of 20 nm.

The notion that mound formation creates long-range correlations in the QW
width fluctuations implies that there should be correlation in the vertical
direction as well. To investigate this issue we have measured the near-field
exciton energies of sample 3, which consists of four different QW grown one
on top of the other. The spectral window of our spectrometer and CCD system
enabled us to measure simultaneously the spectra of three QWs (11, 17 and 25
nm), thus, avoiding any drift that could occur in sequential measurements.
In Fig. 4a we show the fluctuations in the exciton energies of the three QWs
along an 11 $\mu $m line. The near-field exciton lineshapes of the two
narrower wells were non-Gaussian, consisting of multiple peaks. To determine
the characteristic exciton energy, $E_{X}(x,y)$, for each location we
calculated the energy center-of-mass of each spectrum. This procedure
enabled us to get very good accuracy for $E_{X}$. It is evident that there
is an almost perfect correlation between the fluctuation patterns in the
three QWs, indicating that they arise from the same physical origin. It is
seen that this matching in the fluctuation patterns occurs down to
sub-micron length scales. Indeed, we found that the topography of that
sample is characterized by\ shallow and densely packed mounds. We note that
the amplitude of the energy fluctuations increases as the well width
decreases as expected. On the other hand, the bulk GaAs exciton peak, which
is measured simultaneously in the same sample, shows much smaller
fluctuations that are uncorrelated with those of the QWs peaks. This is a
reassuring evidence that the observed fluctuations in the center-of-mass
spectrum are indeed due to well width fluctuations. One can thus conclude
that the process of mound formation indeed creates a strong vertical
correlation between the width fluctuations of QWs grown one on top of the
other.

The existence of this vertical correlation has significant implications for
the understanding of the resonant Rayleigh scattering (RRS) in multiple QWs
samples. It was recently shown that the RRS in multiple QWs is characterized
by temporal oscillations, which are absent in single QW samples \cite%
{Malpuech,Prineas}. These oscillations were explained as resulting from
exciton-polariton effects. However, in presenting this explanation it was
implicitly assumed that full vertical correlation exists between the
disorder in the QWs \cite{Malpuech}. In fact, recent theoretical
calculations have shown that these polaritonic effects depend critically on
the amount of vertical correlation of the disorder between the wells, and
disappear when only moderate correlation exists \cite{Zimmermann}. Our
measurements provide direct evidence to the validity of that assumption, and
substantiate the existence of strong vertical correlation.

We can now quantitatively analyze the relation between the characteristic
amplitude of well-width fluctuations $\Delta L$ and the QW width $L$.
Clearly, $\Delta L$ can be directly obtained from the amplitude of the
exciton energy fluctuation, $\Delta E$, and the mean energy, $E_{QW}$: for a
QW with infinite barriers the relation is simply $\Delta E/E_{QW}=-2\Delta
L/L$, and for finite barriers it can be calculated numerically. In Fig. 4b
we rescale the data of Fig. 4a with the vertical scale being $\Delta L$.
Remarkably, the order of the lines has been changed and the largest
fluctuations in well width are found in the widest well. To further
establish this surprising conclusion we measured the exciton energy of the
four quantum wells along six lines, $11$ $\mu $m long (corresponding to $666$
measurement points), and calculated the standard-deviation, $\sigma $, of $%
\Delta L$ in each QW. In Fig. 5 we plot $\sigma $ as a function of the QW
width $L$. It is clearly seen that $\sigma $ increases monotonically with $L$%
. One can thus conclude that the long-range part of the{\it \ well-width
fluctuations increases with the well-width}. The mechanism that gives rise
to this behavior is illustrated schematically in the inset of Fig. 5. The
drawing shows the AlAs/GaAs and GaAs/AlAs interfaces for wide and narrow
QWs. The coarsening process (discussed above) and the process of smoothening
of the AlAs interface by the GaAs layers give rise to non-uniform growth of
the quantum well. The well is thicker at the mounds minima and thinner at
their maxima. As depicted in the drawing, the amplitude of the well width
fluctuations, which is given by the difference between the wide and narrow
regions, increases with the well width. To verify the correctness of this
model we studied the local relation between $h(x,y)$ and $E_{X}(x,y)$ in all
samples. A detailed examination of the topography and energy images reveals
that at points where $h$ is high $E_{X}$ is also high, and vice versa. In
other words, the well is{\it \ narrow }at places where the surface is high%
{\it \ }and{\it \ wide }where it is low.

In conclusion, we wish to note that the role of surface mounds in
determining the spatial distribution of electrons in MBE-grown samples was
recently realized\ \cite{Willet,Yayon}. Our work shows that this intrinsic
property of MBE growth has far reaching effect on the optical spectrum of
excitons in semiconductor QWs as well.

We are pleased to acknowledge the assistance of V. Zhuk in preparing the
SNOM tips. The research was partially supported by the Israeli Science
Foundation and the Minerva Foundation. A. Esser wishes to acknowledge the
support of LSF grant (HPRI-CT-1999-00069).

\figure Fig. 1: (a), (b) Two-dimensional images of the exciton energy, $%
E_{X}(x,y)$, for samples 1 and 2, respectively.  The energy ranges in (a)
and (b) are 0.2 and 0.5 meV, respectively. (c), (d) The autocorrelation
function $G_{X}(x,y)$ for the range $-10$ $\mu $m $\leq x,y\leq $ $10$ $\mu $%
m, of the exciton images of samples 1 and 2, respectively.

Fig. 2: (a), (b) Surface topography images of samples 1 and 2, respectively.
The full range is 20 nm. (c), (d) The autocorrelation function $G_{h}(x,y)$
for the range $-10$ $\mu $m $\leq x,y\leq $ $10$ $\mu $m, of the topography
images of sample 1 and 2, respectively.

\figure Fig. 3: Cuts through the origin of $G_{X}$ and $G_{h}$ along the
[010] crystallographic orientation of sample 2.

\figure Fig. 4: (a) The exciton energy fluctuations of three QWs grown one
on top of the other along an 11 $\mu $m line. (b) The same line scan as in
(a) translated to well-width fluctuations.

\figure Fig. 5: The standard-deviation of the well-width fluctuations as a
function of the well width. The inset shows a schematic drawing of the
layers structure for narrow and wide QWs. Grey and white areas represent
GaAs and AlGaAs layers, respectively.


\begin{references}
\bibitem{Comment110} The [1\={1}0] direction was ddetermined by the
``bowtie'' defects on the back of the sample.
\bibitem{Comment - dangling bonds} The dangling bonds of these atoms form
dimers that are arranged in rows along the [$1\bar{1}0$] direction, along
which the growth rate is faster.
\end{references}
\end{document}